\documentclass[aps,twocolumn,floatfix,showpacs,amssymb]{revtex4}
\usepackage{graphicx,bm}
\usepackage{hyperref}
\usepackage{amsmath}
\usepackage{mathbbol}
\usepackage{color}

\begin{document}

\title{The three-body pseudo-potential for atoms confined in one dimension}

\author{Ludovic Pricoupenko}

\affiliation
{
Laboratoire de Physique Th\'{e}orique de la Mati\`{e}re Condens\'{e}e, Sorbonne Universit\'{e},  CNRS UMR 7600, F-75005, Paris, France.
}
\date{\today}


\begin{abstract}
Following a strong analogy with two-dimensional physics, the three-body pseudo-potential in one dimension is derived. The Born approximation is then considered in the context of ultracold atoms in a linear harmonic waveguide. In the vicinity of the dimer threshold a direct connection is made between the zero-range potential and the dimensional reduction of the three-body Schr\"{o}dinger equation.  
\end{abstract}

\maketitle

\section{Introduction}

At sufficiently low temperature, ultracold atoms in very elongated harmonic traps can be considered as in a one-dimensional (1D) geometry. For bosons, the idealization of the trap in terms a 1D harmonic waveguide permits one to achieve a mapping with the Lieb-Liniger model \cite{Ols98,Lie63}. In this context, few-body systems in 1D attracted recent interest \cite{Mor05a,Mor05b,Kar09,Mor15}. As a consequence of the Yang-Baxter criterion \cite{Sut07}, the three-body problem occupies also a central place in studying the breakdown of integrability in these systems \cite{Maz08,Tan10,Pet12,Kri16a}. Interestingly,  virtual excitations of pairs of atoms in the transverse modes of the waveguide led to the introduction of a 1D zero-range three-body force~\cite{Maz08,Tan10}. This force has important consequences in the prediction of a 1D dilute liquid state \cite{Pri_Al18,Sek18b}. Moreover, it permits one to predict the existence of an excited trimer state in the vicinity of the dimer threshold \cite{Gui18,Nis18}. This prediction coincides exactly with the one derived directly from the dimensional reduction of  the 3D Schr\"{o}dinger equation in the presence of a 1D waveguide \cite{Pri18a}.

In Ref.~\cite{Nis18} a renormalization procedure is used to cure the divergencies coming from the bare zero-range three-body force. In Ref.~\cite{Gui18} a three-body contact condition is used instead, to implement the zero-range model. These two-last studies lead to exactly the same reduced equation of the three-body problem, referred hereafter as the 1D Skorniakov Ter-Martirosian (STM) equation. Nevertheless the 1D STM equation in Ref.~\cite{Pri18a} differs from the later even though, the same prediction for the excited trimer state is obtained near the dimer threshold. 

In this work, a three-body zero-range pseudo-potential leading to a mathematically well-behaved problem is derived both in configuration (section \ref{sec:Lambda-pot}) and  in momentum space (section \ref{sec:k-representation}). The pseudo-potential has the same form than the two-body ${\Lambda}$-potential in the two-dimensional (2D) space \cite{Ols02}. In section \ref{sec:STM}, 
the STM equation for three-bosons interacting via the  two- and the three-body zero-range forces is derived. This STM equation is identical to the one obtained in Refs.~\cite{Gui18,Nis18}. The ${\Lambda}$ parameter of the pseudo-potential can be any positive real. However the first order Born approximation breaks this ${\Lambda}$-freedom and a specific value of ${\Lambda}$ permits to recover the renormalized strength of the zero-range force in section \ref{sec:Born}. Finally in section \ref{sec:equivalence}, the equivalence between the zero-range 1D model and the dimensional reduction method is obtained in the vicinity of the dimer threshold where the Born approximation is accurate.

\section{Regularized zero-range three-body force}
\label{sec:Lambda-pot}

In this section the 1D three-body contact potential is introduced in complete analogy with the 2D two-body problem. Let us consider three particles of same mass ${m}$ labeled by the index ${i\in (1,2,3)}$. The positions of the three particles are given by the coordinates ${z_i}$. The center of mass of the system is ${C=(z_1+z_2+z_3)/3}$ and the two other Jacobi coordinates are
\begin{equation}
z_{ij}= z_i-z_j\quad ; \quad Z_{ij} = \frac{2}{\sqrt{3}} \left(z_k - \frac{z_i+z_j}{2} \right) ,
\end{equation}
where all the index ${i,j,k}$ are distinct and are a cyclic permutation of the triplet ${(1,2,3)}$. The general form of the zero-range three-body potential can be written as
\begin{equation}
V_3^{\rm pp} \Psi(z_1, z_2, z_ 3) = \frac{\hbar^2}{m} \delta(z_{12}) \delta(z_{23})\psi_3(C) .
\label{eq:V3-contact}
\end{equation}
the function ${\langle C|\psi_3 \rangle}$, is denoted hereafter as the three-body contact. It will will be shown later that it characterizes the singular behavior of the wavefunction at the contact of the three particles. For a given pair ${(ij)}$, one introduces  the hyper-radius ${R=\sqrt{Z_{ij}^2+z_{ij}^2}}$ and the hyper-coordinates ${\mathbf R= z_{ij} \hat{\mathbf e}_z + Z_{ij} \hat{\mathbf e}_Z}$, where ${(\hat{\mathbf e}_z, \hat{\mathbf e}_Z)}$ is an orthonormal basis. The potential in Eq.~\eqref{eq:V3-contact} can be expressed as
\begin{equation}
V_3^{\rm pp} \Psi(z_1, z_2, z_ 3) = \frac{2\hbar^2}{m \sqrt{3}} \delta^{2}(\mathbf R)  \psi_3(C) .
\label{eq:V3-contact-zZ}
\end{equation}
Due to its ${s}$-wave character, the expression of the zero-range potential in Eq.~\eqref{eq:V3-contact-zZ} does not depend on the choice of a specific pair of particles in the definition of the hyper-coordinate ${\mathbf R}$. For the bare three-body zero-range potential used for example in Ref.~\cite{Nis18}, the three-body contact ${\psi_3(C)}$ is just the value of the wavefunction at the contact of the three particles. However, similarly to the two-dimensional Green's function, for a given value of the three-body contact, the two-dimensional delta distribution in Eq.~\eqref{eq:V3-contact-zZ} gives rise to a logarithmic singularity of the wavefunction at ${R=0}$: ${\Psi(z_1, z_2, z_3) \sim \ln \left( R/l \right)  \pi\psi_3(C)/\sqrt{3} + \dots}$ where ${l}$ is a characteristic length of the system. The bare zero-range potential leads thus to a  mathematically not well-behaved model. In practice, the bare potential can be used in the Born approximation, where the wavefunction at first order is regular at the contact \cite{Pri_Al18}. When the bare potential is used non perturbatively as in Ref.~\cite{Nis18}, a renormalizing procedure of the strength is necessary. In order to construct a zero-range pseudo-potential, one has to introduce a regularizing operator such that when applied to the wavefunction considered, it gives a finite value of the three-body contact. For this purpose, one fixes the later characteristics length ${l}$ by imposing a boundary condition at the contact of the three particles as ${R\to 0}$ for all the wavefunctions solution of the Schr\"{o}dinger equation:
\begin{equation}
\Psi(z_1, z_2, z_3) = \frac{\psi_3(C)}{\pi\sqrt{3}} \ln \left( \frac{R}{a_3} \right) + \dots
\label{eq:3B-contact}
\end{equation}
where ${a_3}$ is the 1D three-body scattering length. This boundary condition has been already introduced for the three-body problem in 1D in Ref.~\cite{Gui18}. Equations ~\eqref{eq:V3-contact-zZ} and \eqref{eq:3B-contact} are formally equivalent to the definition of a zero-range force in the 2D two-body problem \cite{Ols02,Pri07}. For instance,  in the free 1D space and in absence of two-body force, one finds from the contact condition in Eq.~\eqref{eq:3B-contact} one trimer at energy ${E_3=-\frac{4e^{-2\gamma}\hbar^2}{ma_3^2}}$, analogous to the dimer in 2D for a zero-range potential.

It is then straightforward to use for the three-body zero-range force, the known expression of the 2D ${\Lambda}$-potential that encapsulates the contact condition of Eq.~\eqref{eq:3B-contact}:
\begin{multline}
V_3^{\rm pp}\Psi (z_1,z_2,z_3)=\frac{-\pi \sqrt{3} \hbar^2}{m \ln\left( e^\gamma \Lambda a_3/2\right) } \delta(z_{12}) \delta(z_{13})\\ \lim_{R \to 0 }\left[ 1 - \ln\left( \frac{e^\gamma \Lambda R}{2}\right)  R \frac{\partial}{\partial R} \right] \Psi (z_1,z_2,z_3) .
\label{eq:pseudo-pot-configuration}
\end{multline}
The parameter ${\Lambda}$ in the pseudo-potential of Eq.~\eqref{eq:pseudo-pot-configuration} is any positive number, i.e. for any value of ${\Lambda}$, the pseudo-potential imposes the contact condition of Eq.~\eqref{eq:3B-contact}. In Eq.~\eqref{eq:pseudo-pot-configuration}, one identifies the ${\Lambda}$-dependent strength
\begin{equation}
g_3(\Lambda) = \frac{-\pi \sqrt{3} \hbar^2}{m \ln\left( e^\gamma \Lambda a_3/2\right) }
\label{eq:g3-Lambda}
\end{equation}
and the regularizing operator
\begin{equation}
\langle C | R_\Lambda |\Psi \rangle =\lim_{R \to 0 }\left[ 1 - \ln\left( \frac{e^\gamma \Lambda R}{2}\right) R \frac{\partial}{\partial R} \right] \Psi (z_1,z_2,z_3) .
\end{equation}
One can verify that for a state satisfying the contact condition in Eq.~\eqref{eq:3B-contact}, one recovers as expected the strength of the zero-range force with ${g_3(\Lambda)\langle C | R_\Lambda |\Psi \rangle=\frac{\hbar^2}{m} \psi_3(C)}$.

\section{Three-body pseudo-potential in the momentum representation}

\label{sec:k-representation}
In what follows, the three-body zero-range pseudo-potential is derived in the momentum representation.  The momentum of the three particles are denoted by ${k_i}$. To avoid any ambiguities with the equations in configuration space, the bra-ket notation will be used below. The analogy with the two-body problem in 2D can be pursued and the derivation follows along the same lines as in Refs.~\cite{Pri10a,Pri11a}. For this purpose, one defines the Jacobi coordinates in the momentum space. The momentum of the center of mass is ${k_C=k_1+k_2+k_3}$. The two other Jacobi coordinates are defined by
\begin{equation}
k_{ij}=\frac{k_i-k_j}{2} \quad ; \quad K_{ij}=\frac{2 k_k - (k_i+k_j)}{3} .
\end{equation}
In what follows, the notations ${k=k_{12}}$ and ${K=K_{12}}$ are used. Similarly to what has been done in the previous section,
one introduces the hyper-momentum 
\begin{equation}
\mathbf Q= k\mathbf u_z+\frac{\sqrt{3}}{2} K \mathbf u_Z .
\end{equation}
The stationary Schr\"{o}dinger equation at energy ${E}$, for a system with only one interaction term given by the three-body force of Eq.~\eqref{eq:V3-contact} is in the momentum space
\begin{equation}
\left(Q^2 + k_C^2/6 + m E/\hbar^2\right) \langle \mathbf Q, k_C | \Psi \rangle = - \langle k_C | \psi_3 \rangle .
\label{eq:Schrodi-Q}
\end{equation}
Equation \eqref{eq:Schrodi-Q} gives the high momentum behavior of the wavefunction for all energies and also in the possible presence of other non singular three-body potential. The three-body contact can be thus defined in the momentum representation by
\begin{equation}
\langle k_C |\psi_3\rangle = - \lim_{Q\to \infty} Q^2 \langle \mathbf Q, k_C |\Psi\rangle .
\end{equation}
Let us consider  the Green's function at the negative energy ${-\hbar^2 \Lambda^2/m}$ in the center of mass frame,  for a vanishing hyper-radius ${(R\to 0)}$:
\begin{equation}
\int \frac{d^2Q}{(2\pi)^2} 
\frac{\exp\left(i\mathbf Q \cdot \mathbf R \right)
}{Q^2+\Lambda^2} = -\frac{1}{2\pi} \ln\left(\frac{e^\gamma R \Lambda}{2} \right)+ \dots
\label{eq:Green-Lambda}
\end{equation}
Equation \eqref{eq:Green-Lambda} can be used for any positive value of the parameter ${\Lambda}$ and permits one to express the three-body contact condition in Eq.~\eqref{eq:3B-contact} as
\begin{equation}
\frac{2}{\sqrt{3}} \int \frac{d^2Q}{(2\pi)^2} \left[ \langle \mathbf Q, k_C | \Psi \rangle + \frac{\langle k_C| \psi_3\rangle }{Q^2+\Lambda^2} \right] = \frac{\hbar^2 \langle k_C| \psi_3\rangle }{m g_3(\Lambda)} .
\label{eq:contact-k-lambda}
\end{equation}
From Eqs.~\eqref{eq:V3-contact} and \eqref{eq:contact-k-lambda}, one can deduce the three-body pseudo-potential in the momentum representation
\begin{equation}
\langle k_1, k_2, k_3 |V_3^{\rm pp}|\Psi \rangle= g_3(\Lambda) \langle k_C| R_\Lambda |\Psi \rangle .
\label{eq:Vpseudo-k}
\end{equation}
where
\begin{equation}
\langle k_C | R_\Lambda |\Psi \rangle =  \frac{2}{\sqrt{3}} \int \frac{d^2Q}{(2\pi)^2} \left[ \langle \mathbf Q, k_C| \Psi\rangle 
+ \frac{\langle k_C |\psi_3\rangle }{Q^2+\Lambda^2} \right] .
\label{eq:regularizing-k}
\end{equation}

\section{STM equation}
\label{sec:STM}

In this section the bound states made of three identical bosons are considered. The three particles interact via the three-body pseudo-potential of Eq.~\eqref{eq:Vpseudo-k}. Moreover,  each pair of particles  ${(ij)}$ interact via the zero-range potential of the Lieb-Liniger model :
\begin{equation}
V(z_{ij}) = - \frac{2 \hbar^2}{ma_2} \delta(z_{ij}),
\end{equation}
where ${a_2}$ is the 1D two-body scattering length. One introduces the two-body contact, which corresponds to the value of the wavefunction at the contact of two particles considered. For the contact of the pair ${(12)}$, one has in the momentum representation :
\begin{equation}
\langle  K,k_C | \psi_2 \rangle = \int \frac{dk}{2\pi} \langle k, K, k_C | \Psi \rangle .
\end{equation}
In the center of mass frame the wavefunction can be factorized as
${\langle k,K, k_C| \Psi \rangle = (2\pi) \delta(k_C) \langle k,K | \phi\rangle}$. 
The  two- and three-body contact  are also factorized as ${\langle K,k_C | \psi_2 \rangle = (2\pi) \delta(k_C) \langle K | S_2 \rangle}$ 
and ${\langle k_C| \psi_3 \rangle = (2\pi) \delta(k_C) S_3}$. 
The three-body Schr\"{o}dinger equation for a bound state of energy ${E=-\frac{\hbar^2q^2}{m}}$  is thus
\begin{multline}
\left(k^2+\frac{3}{4} K^2 +q^2 \right) \langle k,K |\phi\rangle =- S_3 + \frac{2}{a_2}\left(\langle K | S_2 \rangle \right. \\
\left.+ \langle -k-K/2 | S_2 \rangle +\langle k-K/2 | S_2 \rangle \right) .
\label{eq:Schrodi}
\end{multline}
The STM equation follows from Eq.~\eqref{eq:Schrodi} after integration over the relative momentum ${k}$:
\begin{multline}
\left(a_2-\frac{1}{\sqrt{q^2+\frac{3K^2}{4}}}\right) \langle K|S_2\rangle + \frac{a_2 S_3}{\sqrt{4q^2+3K^2}} \\
= 4 \int \frac{dK'}{2\pi} \frac{\langle K'| S_2 \rangle}{K^2+K'\,^2+KK'+q^2} .
\label{eq:STM-1D}
\end{multline}
Injection of Eq.~\eqref{eq:Schrodi} in the contact condition \eqref{eq:contact-k-lambda} gives
\begin{equation}
\frac{3}{a_2}\int \frac{dK}{2\pi} \frac{\langle K | S_2 \rangle}{\sqrt{\frac{3}{4} K^2 +q^2}} =  -\frac{S_3 \ln(q a_3 e^\gamma/2)}{\pi\sqrt{3}} .
\label{eq:closure}
\end{equation}
This last relation permits to close Eq.~\eqref{eq:STM-1D} and to recover the set of equations for the two-body contact obtained in Refs.~\cite{Gui18,Nis18}. Remarkably, an exact implicit equation on the binding wavenumber ${q}$ has been derived from Eqs.~\eqref{eq:STM-1D} and \eqref{eq:closure} in Ref.~\cite{Pri_Al18}.

\section{$\Lambda$-dependent and renormalized strengths}
\label{sec:Born}

In Ref.~\cite{Nis18}, a renormalization of the bare three-body interaction is used for the derivation of the trimer spectrum as a function of the ratio between the two scattering length ${a_3}$ and ${a_2}$. Interestingly, the renormalized strength introduced in Ref.~\cite{Nis18} for a bound state of binding wavenumber ${q}$ coincides with ${g_3(q)}$ \cite{factor}. In the point of view of the pseudo-potential, this corresponds also to the value ${\Lambda=q}$ in the regularizing operator. For this choice, the explicit dependence on the three-body contact in Eq.~\eqref{eq:regularizing-k} is exactly canceled. In the following lines, it is shown that this choice is relevant for the first Born approximation of the zero-range pseudo-potential.

Let us consider a regime where the three-body pseudo-potential gives a small contribution in the eigenenergies. In the center of mass frame, the eigenstate can be decomposed in two parts:
\begin{equation}
|\phi \rangle = |\phi^{(0)} \rangle + |\delta \phi\rangle ,
\end{equation} 
where ${|\phi^{(0)} \rangle}$ is an eigenstate of the three-body problem without the three-body pseudo-potential. The wavefunction ${\phi^{(0)}(z_1,z_2,z_3)}$ is thus regular at the contact of the three-particles and the singular behavior is solely included in the perturbation ${|\delta \phi \rangle}$. In the momentum space, the perturbation ${\langle \mathbf Q | \delta \phi \rangle}$ is  much smaller than ${\langle \mathbf Q | \phi^{(0)}\rangle}$ except in the high momentum limit where 
\begin{equation}
\langle \mathbf Q | \delta \phi \rangle \sim -\frac{S_3}{Q^2+q^2} .
\label{eq:behavior}
\end{equation}
Equation \eqref{eq:behavior} is valid above a given ${Q_0}$ i.e. for ${Q>Q_0 \gg q}$. However, for ${\Lambda=q}$, the contribution in Eq.~\eqref{eq:behavior} is exactly canceled in the regularization and
\begin{equation}
R_{\Lambda=q} |\phi \rangle \sim \frac{2}{\sqrt{3}} \int \frac{d^2Q}{(2\pi)^2} \langle \mathbf Q | \phi^{(0)}\rangle = \phi^{(0)}(R=0).
\label{eq:regularizing-Born}
\end{equation}
Thus, in this perturbative regime the choice ${\Lambda=q}$ permits one to recover the Born approximation of the bare zero-range potential where ${1/|\ln(q a_3)|}$ is the small parameter. 

\section{Three-body problem near the dimer threshold in a 1D waveguide}
\label{sec:equivalence}

In the case of a 1D harmonic atomic waveguide, the effect of virtual excitations of pairs of particles in the direction transverse to the free motion breaks the integrability in the many-body quasi-1D problem. For taking this effect into account, it is necessary to introduce a perturbation to the Lieb-Liniger model. It has been shown that this perturbation can be modeled at the first order Born approximation by a zero-range bare three-body potential with the strength \cite{Maz08}:
\begin{equation}
g_3^{\rm Born} = - \frac{6\hbar^2a_\perp^2}{ma_2^2}\ln\left(\frac{4}{3}\right)  ,
\label{eq:g3-Born}
\end{equation}
where ${a_\perp=\sqrt{2\hbar/(m\omega)}}$ is the characteristic length of the waveguide.

When the 2-body scattering length ${a_2}$ is large and positive, i.e. at the threshold of the dimer of binding wavenumber ${1/a_2}$, the Lieb-Liniger model predicts a shallow trimers (the McGuire trimer) in the 1D waveguide. Virtual excitations of pairs of particles induce a perturbation of this trimer of binding wavenumber ${q=2/a_2}$ \cite{McG64}. This last wavenumber gives the momentum scale of the parameter ${\Lambda}$ chosen when the Born approximation is achieved on the pseudo-potential. At the leading order of logarithmic accuracy, one can then identify the bare strength in Eq.~\eqref{eq:g3-Born} with ${g_3(\Lambda)}$ and ${\Lambda=1/a_2}$. One  finds ${\ln(a_3/a_2) \sim \frac{\pi a_2^2}{4\sqrt{3} \ln(4/3) a_\perp}}$ for ${a_2 \to \infty}$. In this regime, in presence of the three-body potential, in addition to the perturbed McGuire state, another trimer is found near the atom-dimer continuum. Importantly, the asymptotic law near the threshold for the two trimers coincide exactly with the ones found from the dimensional reduction of the 3D Schr\"{o}dinger equation in presence of an harmonic waveguide \cite{Pri18a,Gui18,Nis18}:
\begin{equation}
q_0 \sim \frac{2}{a_2}  +  \frac{4 a_\perp^2}{a_2^3} \ln\left(\frac{4}{3}\right) \, ; \, q_1 \sim  \frac{1}{a_2} + \frac{2 a_\perp^4 }{3 a_2^5} \ln^2 \left(\frac{4}{3}\right) ,
\label{eq:q_trim}
\end{equation}
where ${q_0}$ and ${q_1}$ are the binding wavenumbers.

In the dimensional reduction method of Ref.~\cite{Pri18a}, there is no three-body force in the Hamiltonian and the quasi-1D character of the system is revealed by a summation over the transverse mode in the STM equation: 
\begin{multline}
\left(\frac{1}{\sqrt{\frac{3}{4} K^2 + q^2}} - a_2 \right) \langle K | S_2 \rangle =\\
-2 \int \frac{dK'}{\pi} \sum_{n=0}^\infty
\frac{1}{4^n}\frac{\langle K'| S_2 \rangle}{\frac{4n}{a_\perp^2} + q^2+K^2+K'\,^2+KK'} .
\label{eq:STM-quasi-1D}
\end{multline} 
The momentum ${1/a_\perp}$ plays the role of a cut-off in the integral term of Eq.~\eqref{eq:STM-quasi-1D}. In the limit where the momentum ${K, K'}$ and ${q}$ are much smaller that ${1/a_\perp}$, one has
\begin{equation}
\sum_{n=1}^\infty
\frac{1}{4^n}\frac{1}{\frac{4n}{a_\perp^2} + q^2+K^2+K'\,^2+KK'} 
\sim \ln\left(\frac{4}{3}\right) a_\perp^2 .
\label{eq:approx-STM}
\end{equation}
The approximation in Eq.~\eqref{eq:approx-STM} was the one used for the derivation of the asymptotic law for the two trimers in Eq.~\eqref{eq:q_trim}. In what follows it is shown that this approximation is equivalent to the Born approximation of Eq.~\eqref{eq:g3-Born} performed with the zero-range pseudo-potential. 

For that purpose, one considers formally the effective three-body potential ${V_3}$ whhich gives Eq.~\eqref{eq:STM-quasi-1D}. The Hamiltonian is
\begin{equation}
H=-\frac{\hbar^2}{2m}\sum_{i=1}^3 \frac{\partial^2}{\partial z_i^2} - \frac{2 \hbar^2}{ma_2} 
\sum_{i<j\le 3} \delta(z_{ij}) + V_3 .
\end{equation}
The absence of singular behavior of the eigenstates of Eq.~\eqref{eq:STM-quasi-1D} shows that necessarily the potential ${V_3}$ is non-local in the configuration space. One can remark that the regime where the two-body contact ${\langle K |S_2 \rangle}$ is negligible for a momentum ${K}$ of the order of ${1/a_\perp}$, corresponds to the regime where the zero-range three-body force can be treated in the Born approximation. Then using the approximation Eq.~\eqref{eq:approx-STM}, one finds an energy dependent potential valid in the Born approximation. For an hyper-momentum ${Q \lesssim 1/a_\perp}$, one can write in the center of mass frame
\begin{multline}
\langle k,K |V_3|\phi \rangle=  -\frac{\hbar^2}{m} \pi \ln(4/3) (k^2+\frac{3}{4}K^2+q^2) \frac{a_\perp^3}{a_2} \\
\times \int_< \frac{dk'dK'}{(2\pi)^2} \langle k',K' | \phi \rangle 
\label{eq:V3-non-local}
\end{multline}
and ${\langle k, K |V_3|\phi \rangle \sim 0}$ for ${Q \gtrsim 1/a_\perp}$. In Eq.~\eqref{eq:V3-non-local}, ${\int_<}$ means an integration in the domain where the hyper-momentum ${Q'=\sqrt{k'\,^2+\frac{3}{4}K'\,^2}}$ is smaller than ${1/a_\perp}$. Multiplying  Eq.~\eqref{eq:V3-non-local} by ${\langle \phi|k, K\rangle}$ and assuming that ${V_3}$ is a small perturbation, one can use the relation 
\begin{multline}
\langle  \phi|\mathcal H_0-E|k, K \rangle  \sim  \frac{2\hbar^2}{ma_2}\left(\langle  S_2 |K \rangle \right. \\
\left.+ \langle S_2 | -k-K/2 \rangle +\langle S_2 |k-K/2  \rangle \right) .
\label{eq:bra_Hamiltonian}
\end{multline}
The integration of the resulting equation on the hyper-momentum in the domain ${Q < 1/a_\perp}$ gives the expectation value of the effective three-body potential in the center of mass frame :
\begin{equation}
\langle \phi |V_3| \phi \rangle \sim  - \frac{6\hbar^2 a_\perp^2 }{ma_2^2} \ln\left(\frac{4}{3}\right) |\phi(R=0)|^2.
\end{equation}
One thus recovers the same strength of the three-body potential as the one of the zero-range model in the Born approximation in Eq.~\eqref{eq:g3-Born}.

\section{Conclusion}

The three-body ${\Lambda}$-potential introduced in this paper leads to a mathematically well-behaved Schr\"{o}dinger equation, avoiding thus a renormalization procedure of a bare zero-range force. This pseudo-potential is used in the context of atoms in a 1D waveguide where the virtual excitations in the transverse modes give rise to an effective zero-range three-body force in addition to the usual two-body force of the Lieb-Liniger model. For the three-body problem, in the limiting case of a large two-body scattering length, a three-body potential is obtained from the STM equation derived directly from the 3D Schr\"{o}dinger equation in the presence of a 1D waveguide. This potential treated at the Born approximation explains the equivalence found near the dimer threshold, between the zero-range potential approach and the direct method of Ref.~\cite{Pri18a}.


\begin{thebibliography}{99}
\bibitem{factor} The definition adopted here for the three-body scattering length ${a_3}$ is the same as the one in Ref.~\cite{Gui18} but differs by a factor ${e^\gamma/2}$ from the one used in Ref.~\cite{Nis18}.
\end{thebibliography}
\end{document}